# T2 mapping at 0.55 T using Ultra-Fast Spin Echo MRI

Margaux Roulet MS[1,,+], Hamza Kebiri PhD[1,2], Busra Bulut MS[1,2], Vladyslav Zalevskyi MS[1,2], Thomas Sanchez PhD[1,2], Jean-Baptiste Ledoux[1,2], Vincent Dunet MD[1], Mériam Koob[1], Tom Hilbert PhD[3], Tobias Kober PhD[4], Erick J Canales-Rodriguez PhD[1,2,5,6,7], Meritxell Bach Cuadra PhD[1,2,+]

[1]Department of Medical Radiology, Lausanne University Hospital (CHUV) and University of Lausanne (UNIL), Lausanne, Switzerland

[2]CIBM Center for Biomedical Imaging, Switzerland

[3]Advanced Clinical Imaging Technology, Siemens Healthineers International AG, Lausanne, Switzerland

[4]Research & Clinical Translation, Magnetic Resonance, Siemens Healthineers AG, Erlangen, Germany

[5]Signal Processing Laboratory 5 (LTS5), École Polytechnique Fédérale de Lausanne (EPFL), Lausanne, Switzerland

[6]Department of Signal Theory, Networking and Communications, University of Granada, Granada, Spain

[7]Andalusian Research Institute in Data Science and Computational Intelligence (DaSCI), University of Granada, Spain

**Corresponding Author Info:**

[+]Margaux Roulet & Dr. Meritxell Bach Cuadra

Address: Department of Medical Radiology, Lausanne University Hospital (CHUV) and University of Lausanne (UNIL), Lausanne, Switzerland, Centre de Recherche en Radiologie PET03/02/210, CHUV, Rue du Bugnon 46, CH-1011 Lausanne, Switzerland

Phone: +41-21-314-67-75

e-mail: margaux.roulet@unil.ch, meritxell.bachcuadra@unil.ch

**Acknowledgment:**

We thank Dr Ruud van Heeswijk for having provided access to the NIST phantom. We acknowledge the Leenaards and Jeantet Foundations as well as CIBM Center for Biomedical Imaging, a Swiss research center of excellence founded and supported by CHUV, UNIL, EPFL, UNIGE and HUG.

**Author contribution:**

M. Roulet: Conceptualization, Investigation, Methodology, Software, Visualization, Data Acquisition, Writing – Original Draft Preparation. H. Kebiri: Methodology, Data Acquisition, Review & Editing, Supervision. B. Bulut: Writing, Review & Editing, V. Zalevskyi: Software, Methodology, Review & Editing. T.Sanchez: Methodology, Review & Editing. J-B Ledoux: Data Acquisition, Review & Editing. V. Dunet: Data Acquisition, Review & Editing. M. Koob: Data Acquisition, Review & Editing. T. Hilbert: Review & Editing. T. Kober: Methodology, Review & Editing. E. J C Rodriguez: Methodology, Review & Editing. M. Bach Cuadra: Funding Acquisition, Supervision, Conceptualization, Methodology, Review & Editing. The authors acknowledge the use of AI writing assistance tools (DeepL, ChatGPT) to assist with spell checking, grammar refinement, and language clarity improvements. After using this tool/service, the authors reviewed and edited the content as needed and take full responsibility for the content of the publication.

**Grant Support:** This research was funded by the Swiss National Science Foundation (215641), ERA-NET Neuron MULTI-FACT project (SNSF 31NE30_203977);

## Abstract

**Background:** Low field T2 mapping magnetic resonance imaging has the potential to democratize neuropediatric imaging by enhancing accessibility thanks to cost-effectiveness and by providing quantitative tissue biomarkers of the developing brain.

**Purpose:** To conduct a high resolution T2 mapping feasibility study based on SS FSE sequence at 0.55 T on a control healthy cohort.

**Study Type:** Prospective single-center study.

**Population:** In vivo: Ten healthy adult subjects (18–43 years, females/males: 5/5). In vitro: NIST Phantom.

**Field strength/sequence:** Ultra-fast spin-echo multi-echo sequence at 0.55T and 1.5 T.

**Assessment:** Feasibility is assessed in vitro using the NIST Phantom, comparing T2 relaxation times to spectrometer references measured at 0.55 T. Acquisition and T2 fitting protocol are optimized in vitro for in vivo analysis. In vivo repeatability is assessed with atlas-based evaluation on White Matter (WM) and Cortical Grey Matter (GM), with ROIs delineated within each tissue type. Coefficient of variation inter-run/session/subject was assessed.

**Statistical Tests:** A Wilcoxon signed-rank test with Bonferroni correction was used to assess statistical differences in coefficient of variation across run, session, and subject ($\alpha$=0.05/n$_{ROI}$). Pearson correlation coefficients were calculated for T2 times.

**Result:** We demonstrate feasibility in vitro using mono-exponential fitting under Gaussian-Rician noise assumptions with deviation below 12% from spectrometer references. In vivo, the inter-subject CoV was 5.2% for WM and 17.7% for GM, showing no significant difference compared to 1.5 T. T2 times were 118 ms and 188 ms at 0.55 T for WM and GM, respectively, with an acquisition time of 16.5 minutes.

**Conclusion:** We present a rapid, robust high resolution T2 mapping protocol at 0.55 T for HASTE MRI, using Gaussian noise-based T2 fitting and report for the first time T2 times for healthy adult brains at 0.55 T, demonstrating efficiency and reliability.

## <u>Introduction</u>

Magnetic Resonance Imaging (MRI) is an essential diagnostic imaging, particularly in the field of pediatric and perinatal neuroimaging (1–3). Additionally, its non-invasive nature and excellent soft tissue contrast make it a powerful technique for assessing in-vivo brain development (4–6) and detecting structural abnormalities in early stages (6, 7). Among the various MRI contrasts, T2-weighted imaging is particularly valuable due to its high sensitivity to microstructural changes that reflect white matter (WM) and grey matter (GM) tissue maturation (1, 4, 5, 8). However, significant variability in acquisition protocols across centers makes it challenging to gather multicentric data.

Quantitative T2 mapping offers a potential solution by providing consistent relaxometry values despite the scanner differences and acquisition settings (8–11). It enables reproducible, field-strength-dependent measurements of voxel-wise T2 relaxation times, improving the specificity and sensitivity for identifying pathologies and congenital disorders through quantitative comparisons with normative healthy pediatric cohorts (10, 11). Despite its promise, quantitative relaxometry is rarely implemented in clinical practice in fetal MRI, due to the long acquisition times required and the susceptibility to motion artifacts from fetal movement and maternal respiration (12). Clinical fetal MRI protocols instead rely on low-resolution T2-weighted single-shot fast spin echo (SS-FSE) sequences to freeze motion within each slice. High-resolution (HR) volumes can then be reconstructed using interpolation or super-resolution (SR) techniques (11, 13–15), but these methods are rarely integrated into quantitative pipelines.

Recent advances have renewed interest in low- and mid-field MRI systems (e.g., 0.064T, 0.55T) (16–18), which, despite their inherently lower signal-to-noise ratio (SNR), offer several advantages. These include reduced distortions (19), prolonged intrinsic T2 relaxation times, and a larger dynamic range of T2 values (15–18). Such properties enhance the visualization of novel tissue contrasts, particularly in regions with low T2 values like deep GM (15). In addition to these technical benefits, low-field MRI systems are cost-effective, presenting a viable solution for expanding access to MRI diagnostics beyond tertiary care centers in high-resource settings (16, 20). Indeed, fetal MRI is particularly relevant in the context of brain development research in conditions like malnutrition and pre-natal stress, which are disproportionately prevalent in underserved populations. The application of quantitative T2 mapping within low-cost, low-field MRI platforms may enable researchers and clinicians in low- and middle-income countries to assess brain maturation that is comparable across diverse populations (20).

## Related Works

Pioneering works on fetal low-field MRI has focused on T2* mapping, typically using gradient-echo single-shot EPI sequences (15, 21). These approaches are inherently rapid and three-dimensional, avoiding the need for HR reconstruction. A recent study employed T2* mapping at 0.55T and demonstrated significant potential for placental and fetal brain evaluation as an indirect measure of oxygenation, with mean T2* values exhibiting a linear decline over the course of gestation (15).

In contrast, T2 relaxometry, which offers a direct quantitative measure of spin-spin interactions, has been explored in only a single study at 1.5T for fetal brain imaging (11). Early simulation and phantom studies revealed an overestimation of T2 values with SS-FSE sequences, especially when using mono-exponential decay models, relative to spectrometer-derived ground truth (13). To mitigate this bias, regularization techniques, often used to solve ill-posed problems, have been applied to introduce prior knowledge about the acquired signal (22–25). Other approaches take inspiration from MR fingerprinting (26) and involve adapting multi-echo T2w sequences for simulations (27), using either Bloch equations or the extended phase graph (EPG) formalism (28–30). A notable contribution includes a dictionary-based fitting method using EPG, combined with super-resolution volume reconstruction (SVRTK) (11). This approach improved accuracy in fetal T2 mapping using SS-FSE sequences at 1.5T.

Despite increased interest in low-field MRI, T2 mapping of ultra-fast spin echo sequences at low field remains underexplored, with no established in vivo reference values. To fill this gap, validation using physical phantoms, which provide access to spectrometer-derived ground truths, is essential. Such phantoms allow for reproducible in vitro assessment of acquisition and reconstruction methods (31–33). Moreover, applying these protocols to well-characterized populations, such as healthy adults, provides an important benchmark. Recent studies using standard spin echo sequences have evaluated T1 and T2 measurements in both phantoms and healthy volunteers at 0.55T, showing statistically comparable performance across prototype and commercial systems (34) .

## Contributions

*Work in preparation for submission at JMRI*

This work demonstrates the feasibility of high-resolution T2 mapping at 0.55 T using SS-FSE sequences, validated in a controlled setting with healthy adult volunteers. We present the first in vivo evidence of its applicability at 0.55 T, alongside a performance comparison with 1.5 T HASTE acquisitions. It also provides initial data on T2 relaxation times of brain tissue in healthy adult volunteers at 0.55T. The primary aim was to establish and validate the technique under controlled conditions thus independently of reconstruction quality or motion-correction constraints inherent to fetal data. We discuss the benefits and potential strategies for extending this technique to fetal and newborn cohorts to obtain low-field quantitative MRI measurements for developing brains.

## Materials and Methods

This study was approved by the ethics committee of the Canton of Vaud, Switzerland (project number: 2024-04995).

### MRI acquisition and pre-processing

MR acquisitions are conducted both at 0.55 T and 1.5 T (MAGNETOM Free.Max and Sola, Siemens Healthineers, Forchheim, Germany). Clinical T2w series using an ultra-fast Half-Fourier Acquisition Single shot Turbo spin Echo (HASTE) sequence and comprising 2D slices in the three orthogonal orientations are acquired with a repetition time (TR) of 2500 ms and at a resolution of $1.2 \times 1.2 \times 4.5$ mm3 (ESP = 8.8 ms at low field and ESP = 3.9 ms at high field, excitation/refocusing pulse flip angles of 90°/180°, no interslice gap). To mimic the clinical fetal protocol, a limited field-of-view (FOV, 360×360 mm2) is imaged. GRAPPA acceleration is used (Number of reference line=2, Acceleration=42). Acquisitions are repeated with nine echo times (TE) uniformly sampled over the range of 114 ms to 299 ms ($\Delta$TE $\approx$ 15 ms). Each series contains 40 slices and is acquired in 1.8 minutes. The total acquisition time of three different orthogonal volumes per TE, ranges from 16.5 minutes (three TEs) to 49.5 minutes (nine TEs). Acquisitions are performed both in vitro on the NIST phantom (35) using both head coil and soft body coil, and in vivo on adult brains using head coil (18 channels Body Array combined with 32 channels spine coil at 1.5 T and 12 channels Contour Medium Coil combined with 9 channels spine coil at 0.55 T). For each TE, the three orthogonal low resolution (LR) series are resampled to an isotropic voxel size of 1 mm$^3$. Subsequently, rigid co-registration and integration into a single high-resolution (HR) volume using trilinear interpolation is performed followed by total-variation denoising (36). The output of the processing pipeline is a HR T2w volume (for each TE), which will be used for the construction of T2 relaxation maps as summarized in Fig. 1.

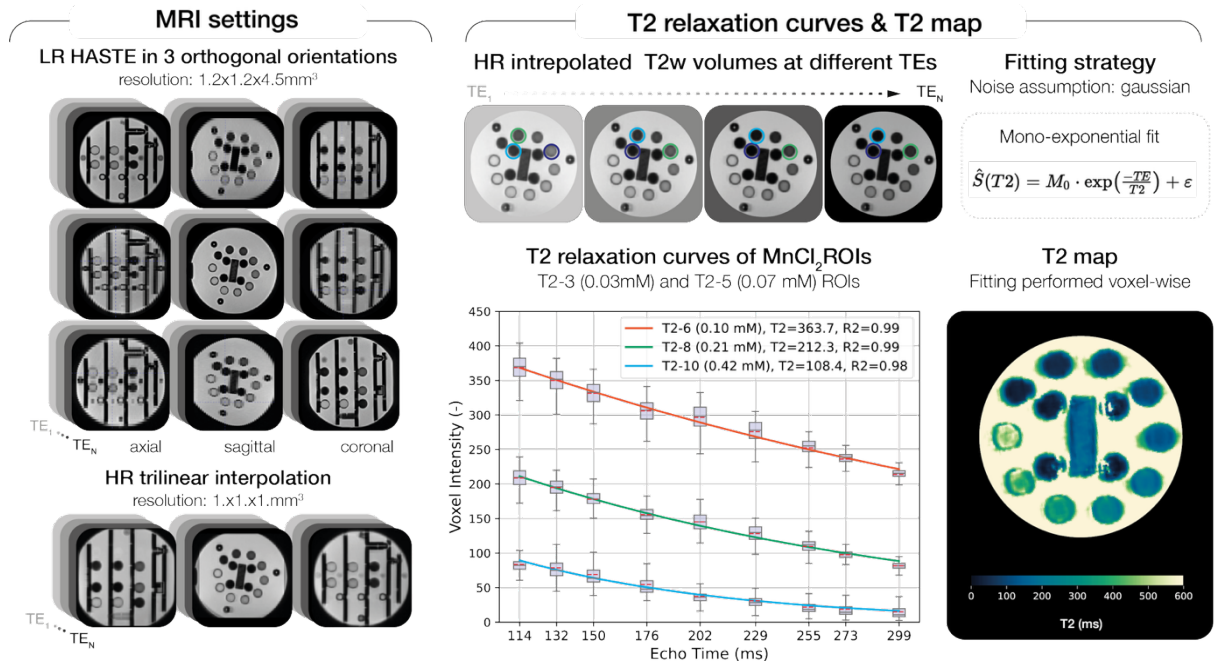

**Figure 1: In-vitro proof-of-concept T2 mapping and evaluation framework at 0.55 T.** Ultra-fast spin-echo multi-echo low-resolution (LR) HASTE acquisitions are performed in three orthogonal orientations (axial, sagittal, coronal), followed by trilinear interpolation for each echo time series. High-resolution T2-weighted volumes are evaluated for feasibility in T2 relaxation assessment across three selected ROIs of the NIST Phantom (MnCl2 array: T2-7, T2-9, T2-11). T2 maps are generated through voxel-wise mono-exponential decay fitting.

## T2 mapping

T2-weighted signal intensity follows an exponential decay governed by the tissue-specific transverse relaxation time (T2). While a multi-exponential model can theoretically provide a more accurate fit—particularly in accounting for compartments with shorter T2 values (10–20 ms)—this approach is not feasible in fetal MRI due to the prolonged acquisition times required. Moreover, given that the TEs of interest are much longer than the T2 of fast-decaying components, their signal contribution is effectively negligible (13). Still, in partial volume voxels, eg, at interface of WM/GM or GM/CSF, the estimated T2 represents a weighted average of the contributing tissues. Therefore, a mono-exponential fit is adopted to describe the signal intensity A using:

$$A = M_0 e^{-TE/T2} + \epsilon \qquad (1)$$

, where TE is the effective echo time chosen, T2 is the relaxation time, M0 is the transverse magnetization. This model assumes a Gaussian noise distribution $\epsilon \sim \mathcal{N}(0, \sigma^2)$. However, given the lower signal-to-noise ratio (SNR) at 0.55 T field strengths, it may be more appropriate to account for the Rician distribution of noise inherent to complex MRI signals (37, 38). In the presence of Rician noise, the probability distribution of the measured signal intensity is given by the Rice density:

$$p(S) = \frac{S}{\sigma^2} e^{-\frac{S^2 + A^2}{\sigma^2}} I_0 \left( \frac{S * A}{\sigma^2} \right) \qquad (2)$$

where $A$ denotes the signal intensity in the absence of noise, $I_0$ is the modified zeroth order Bessel function of the first kind and $\sigma$ denotes the standard deviation of the Gaussian noise in the real and the imaginary images. In the case the noise follows a purely Rician distribution, T2 fitting can be done by minimizing the negative log-likelihood of the Rician noise distribution (39).

A Gaussian approximation of the Rician distribution is obtained if A = 0 when only noise is present. It was demonstrated by (40) that for image regions with large signal intensities the noise distribution can be considered as a Gaussian distribution with variance $\sigma^2$ and mean $\sqrt{A^2 + \sigma^2}$. Therefore, we will also explore T2 fitting using the following Gaussian-Rician approximation of the noise:

$$S = \sqrt{(M_0 e^{-TE/T2})^2 + \sigma^2} \qquad (3)$$

T2 maps are generated by voxel-wise fitting of the signal intensity to models (1) or (3), solving a minimization problem using the L-BFGS-B optimization algorithm. Optimization parameters are available in Table S1.

## In vitro Feasibility Study

To our knowledge, no ground truth nor references values are available for T2 values at 0.55 T for brain tissues. In vitro studies can help though as a first step to assess the feasibility and accuracy of our approach, prior to in vivo experiments. Therefore, feasibility study is first done with a quantitative evaluation using the NIST phantom (35). We restrict the study to regions of interest (ROIs) with a high T1/T2 ratio (on the $MnCl_2$ plate) that mimic fetal tissue properties and for which spectrometry ground truth estimation is available at 0.55 T (41). Our study will therefore focus on the samples labelled as T2-5 (0.07 mM), T2-7 (0.1 mM), T2-9 (0.28 mM) and T2-11 (0.56 mM), with respective spectrometric reference T2 values of 284 ms, 167 ms, 80 ms and 40 ms at 0.55 T. We analyze T2 mean, standard deviation and mean absolute percentage error relative to the spectrometer reference. We do an ablation study and explore the impact of different T2 fitting under (a) different noise assumptions: Gaussian, Gaussian-Rician, Rician; (b) distinct coil: head coil and soft body coil; (c) distinct number of echo time (TE): 3,5,9; and (d) distinct TE coverages: ranging in 114-299ms, 114-202ms, 202-299ms.

The findings of this in vitro experiment will assess the suitability of the acquisition approach and enable optimization of the T2 mapping protocol for in vivo analyses detailed here after.

### In vivo Repeatability Study

Ten healthy subjects (five females, five males, mean age = 27, age range = [18,43]) were enrolled in the study. Written informed consent was obtained from each participant enrolled.

Using the optimal T2 mapping strategy and echo time (TE) coverage established in vitro, we acquired images at 0.55 T and 1.5 T and generated high-resolution three-dimensional (3D) T2 maps for each adult subject. For each subject, mimicking fetal clinical protocols, T2-weighted orthogonal HASTE sequences are acquired at three TEs (114 ms, 202 ms, and 299 ms), rigid registration and trilinear interpolation was performed as described Section MRI acquisition and preprocessing, and T2 fitting was performed under the assumption of Gaussian noise. WM and GM tissues were segmented using SynthSeg (42). T2 maps were then registered to the MNI152 standard space using FSL FLIRT affine registration (43). Atlas-based regions of interest (ROIs) within WM and GM were delineated using the ICBM-DTI-81 (44) and Harvard-Oxford Cortical (45) atlases. 27 and 41 ROIs were selected for WM and GM tissue type as detailed in Tables S4 & S5.

To study the robustness of the in vivo acquisition and T2 fit protocol, a repeatability study is carried out. Recognizing that variability can come from different sources (46), we examined three factors, similar to a previous study carried out at 1.5 T (47). For WM and GM atlas-based ROIs, we studied inter-run, inter-session and inter-subject variability: (i) inter-run variability was assessed using scans obtained during the same session on the same scanner for the same subject (N=2); (ii) inter-session variability was determined from scans acquired across different sessions of the same subject on the same scanner (N=2); and finally, (iii) inter-subject variability was determined from scans acquired across different subjects on the same scanner (N=10). Inter-subject variability at 0.55 T was also compared to variability at 1.5 T.

Variability of T2 values within each tissue (WM and GM) is assessed using (a) the coefficient of variation (CoV), which is calculated for each tissue region-of-interest (ROI) as the standard deviation (SD) of the scalar measurements divided by the ROI mean, multiplied by 100. Intuitively, the CoV reflects the fraction of the mean scalar measurement attributable to variability. Thus, a higher CoV signifies greater variability. Pairwise comparisons of variability due to run, session, and subject effects were performed using a Wilcoxon signed-rank test with Bonferroni correction (significance level: $\alpha$=0.05/$n_{ROI}$).

## Results

### Protocol optimization in-vitro

Voxel-wise T2 maps of the NIST phantom using the HASTE sequence demonstrated consistent relaxation decay patterns but revealed systematic overestimation of T2 times (Figure 2 and Table 1). Qualitatively, the Gaussian noise model produced T2 maps with fewer visible artifacts and more uniform voxel-wise estimates, whereas the Gaussian-Rician and Rician fits reduced T2 overestimation but introduced inhomogeneity and instability (Figure 2).

Comparisons with spectrometer and single-echo references revealed that SE acquisitions fitted with the Gaussian model yielded 1/T2 slope deviations of only 2.8% from reference values, while HASTE acquisitions produced larger deviations up to 46%. Gaussian-Rician fitting partially mitigated this bias, achieving slope deviations below 12.7%, but with increased standard deviations and visibly noisier maps.

Ablation study on echo times to optimize the acquisition protocol found that using three TEs is sufficient to fit the T2 relaxation decay (Figure S2, Table S5). Narrowing the TE coverage from 114-299 ms to 144–202 ms, reduced the average T2 overestimation but limited applicability for tissues with longer T2 values.

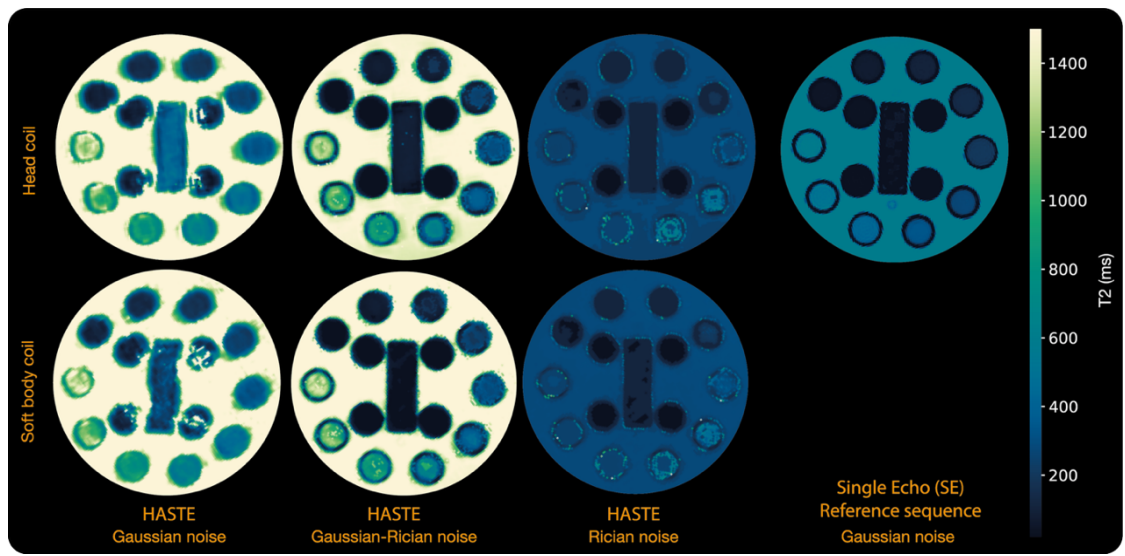

**Figure 2: In vitro T2 maps of the NIST Phantom under three noise models: Gaussian, Gaussian-Rician, and Rician.** Comparisons are shown between soft body and head coils when T2 maps with Gaussian and Gaussian-Rician noise assumptions yield comparable quantitative results, though the Gaussian model produces smoother maps. Under Rician noise, T2 maps show deviations with sharper transitions and fail to predict T2 values above 300 ms accurately. Maps can be compared to the Single Echo reference sequence.

**Table 1**: In vitro ablation study on noise distribution assumption (head coil – body coil).

| | ROI | mM | SP Ref[*] | SE Ref[**] | HASTE Head coil | | | HASTE Soft body coil | | |
|---|---|---|---|---|---|---|---|---|---|---|
| fit | | | - | Gaussian | Gaussian | Gaussian-Rician | Rician | Gaussian | Gaussian-Rician | Rician |
| | T2-3 | 0.031 | 593.85 | 431 ± 11 | 841 ± 53 | 850 ± 70 | 259 ± 53 | 834 ± 53 | 880 ± 118 | 252 ± 1 |
| | T2-4 | 0.047 | 416.01 | 338 ± 10 | 652 ± 52 | 552 ± 131 | **476 ± 185** | 644 ± 54 | 637 ± 129 | **417 ± 190** |
| | T2-5 | 0.069 | 283.87 | 273 ± 5 | 497 ± 39 | **357 ± 57** | 436 ± 104 | 492 ± 46 | **383 ± 55** | 437 ± 103 |
| T2 (ms) | T2-6 | 0.101 | 221.27 | 202 ± 5 | 352 ± 22 | 318 ± 43 | **292 ± 52** | 338 ± 28 | 311 ± 28 | **306 ± 45** |
| | T2-7 | 0.145 | 166.76 | 145 ± 2 | 272 ± 22 | 232 ± 62 | **135 ± 35** | 260 ± 25 | 247 ± 39 | **187 ± 39** |
| | T2-8 | 0.207 | 121.51 | 102 ± 2 | 211 ± 20 | **118 ± 43** | 110 ± 0 | 189 ± 19 | 143 ± 41 | **110 ± 0** |
| | T2-9 | 0.296 | 80.43 | 73 ± 1 | 154 ± 19 | **76 ± 4** | 110 ± 0 | 150 ± 24 | **77 ± 10** | 110 ± 0 |
| | T2-10 | 0.421 | 53.23 | 48 ± 1 | 111 ± 38 | 12 ± 10 | 91 ± 32 | 128 ± 40 | 12 ± 11 | 71 ± 43 |
| | T2-11 | 0.599 | 41.08 | 37 ± 1 | 119 ± 71 | 10 ± 0 | 25 ± 28 | 136 ± 94 | 11 ± 0 | 20 ± 21 |
| | 1/T2 regression slope[***] (ms⁻¹) | | 39.2 | 40.3 | 19.9 | **44.2** | 29.2 | 21.1 | **42.2** | 27.9 |
| | % err | | - | 2.8 | -49.2 | **12.7** | -25.5 | -46.6 | **7.6** | 28.8 |

*[*]Ref.= spectrometer references value measured at 0.55 T (34). [**]A single slice SE reference sequence. A total of 27 TEs were used spanning from 12 to 428 ms, no prescan, head coil was used. [***]Relaxation rates slopes are computed excluding T2-10 and T2-11.*

Taken together, these findings indicate a clear stability-accuracy trade-off, and despite larger deviation from reference 1/T2 slopes, the Gaussian noise model remains the most stable choice for voxel-wise T2 mapping across a wide range of T2 values. Based on these results, the final protocol uses a Gaussian fitting model with three echo times ([114, 202, 299] ms) and has a total acquisition time of 16.5-minutes, with the caveat that potential T2 overestimation in vivo should be accounted for in further analysis.

### In vivo feasibility

In vivo voxel-wise T2 maps revealed consistent relaxation times (Fig. 3), with average T2 times of 118 ms for WM and 188 ms for GM at 0.55 T across subjects (Table S6). The T2 times distribution for WM was consistent across subjects too, with narrow tails and a centered mean (Fig. 4, left). In contrast, the T2 distribution for GM was broader, with extended upper tails. At 1.5 T, average T2 times were 111 ms for WM and 171 ms for GM, aligning with previous studies at 1.5 T for WM but slightly overestimated for GM (34).

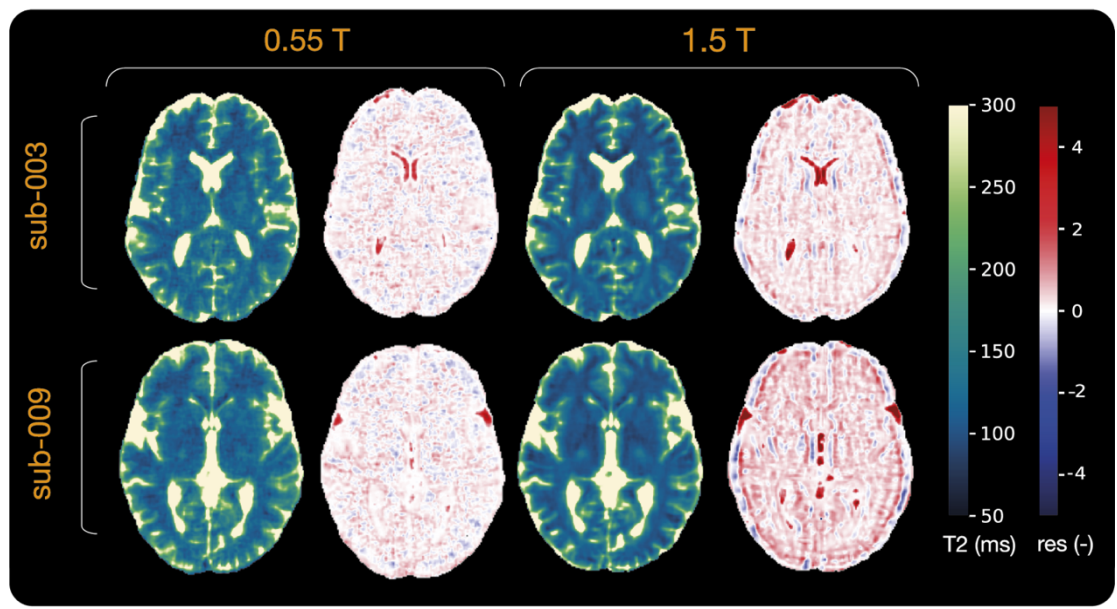

**Figure 3: In vivo T2 maps and residuals for two subjects at 0.55 T and 1.5 T.** The evaluation framework effectively mitigates SNR reduction in T2 maps at 0.55 T. As expected, T2 values are higher at 0.55 T than at 1.5 T. Residual maps at 1.5 T and 0.55 T remain unbiased across white and grey matter. Absolute residual error is greater at 1.5 than at 0.55 T.

Partial volume effects are less pronounced for GM at 0.55 T than at 1.5 T (Fig. 3), particularly at the interface with cerebrospinal fluid, as highlighted by the residual maps: at 0.55 T, the residual bias is relatively uniform, without clear tissue-specific patterns, while at 1.5 T, distinct residual structures are visible for WM and GM, with higher absolute residuals amplitude overall.

## In vivo repeatability

Figure 4 illustrates the inter-run, inter-session, and inter-subject variability for both GM and WM. The CoV for the same subject does not increase significantly inter-session, rising to 1.4% for GM and 0.8% for WM, compared to inter-run values of 1.5% for GM and 0.5% for WM (p = 0.69 for GM and p = 0.005 for WM, Table S2 and S3). Inter-subject variability reached a CoV of 5.24% for WM and 17.7% for GM at 0.55 T, comparable to measurements at 1.5 T, where CoV values were 5.00% for WM and 12.5% for GM. A Wilcoxon signed-rank test, with significance level adjusted by Bonferroni correction ($\alpha$=0.05/$n_{ROI}$), showed no significant difference in inter-subject CoV between 0.55 T and 1.5 T (p = 0.95 for WM and p = 0.06 for GM, Table S2 and S3). Regression analysis comparing two runs within the same session, or two different sessions demonstrated a strong correlation with Pearson correlation coefficients of respectively 0.99 and 0.97 for WM, and 0.96 and 0.94 for GM. While inter-subject correlation was milder for WM, with coefficients of 0.7 for WM and 0.91 for GM, it remained coherent across both tissue types. Finally, comparing 0.55 T and 1.5 T, Pearson correlation coefficients was 0.89 for WM and 0.63 for GM.

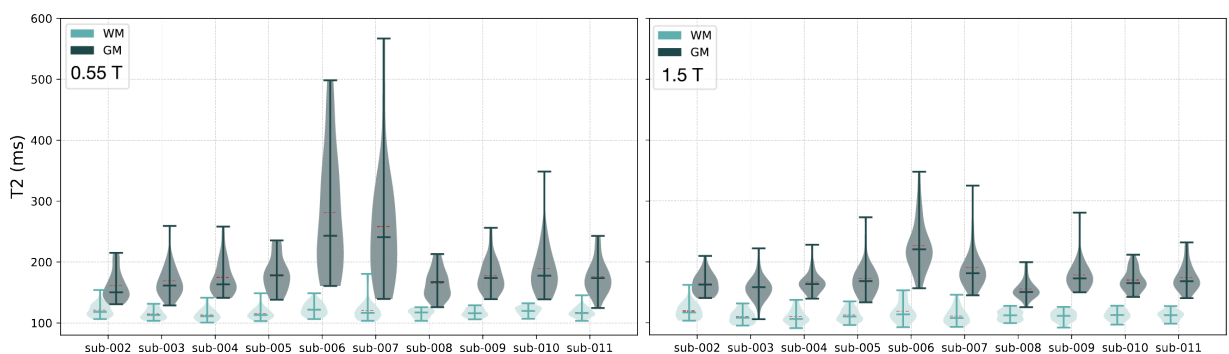

**Figure 4: T2 distribution per subject in white and grey matter at 0.55 T and 1.5 T.** Consistent T2 times are observed across both field strengths. White matter shows a more compact distribution compared to grey matter, where larger dispersion—due to partial volume effects—is evident in some subjects. White matter dispersion remains similar across field strengths, while grey matter dispersion increases for specific subjects at 0.55 T.

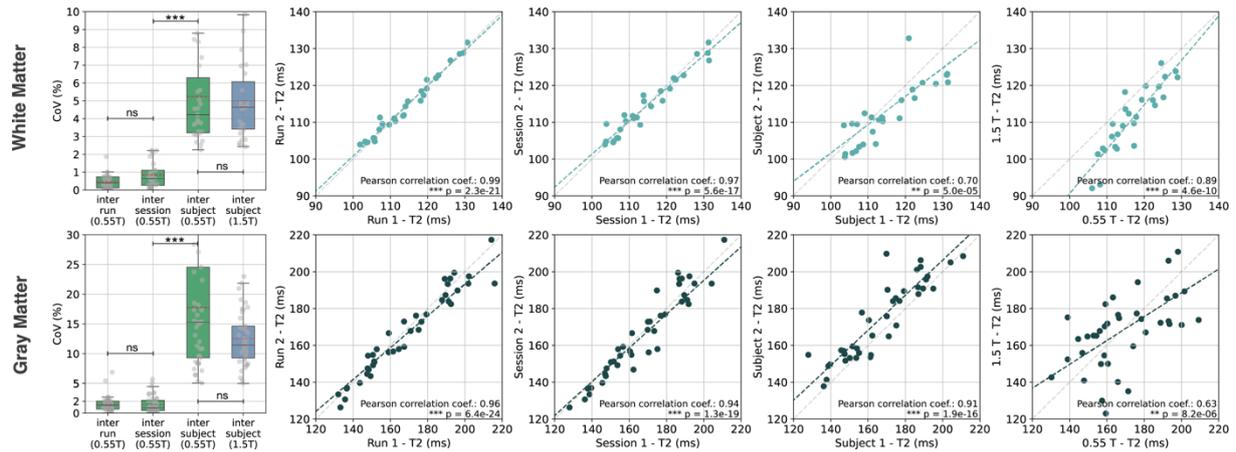

**Figure 5: Repeatability study across runs, sessions, and subjects**. Each scatter plot shows ROI data from (top) the ICBM-DTI-81 white matter atlas and (bottom) the Harvard-Oxford cortical grey matter atlas. **Bar plot**, Coefficient of Variation (CoV %) reflects inter-run, inter-session, and inter-subject variability at 0.55 T, and inter-subject variability at 1.5 T. **Regression analysis**, with x- and y-axes representing T2 values across two runs, sessions, subjects, or field strength.

## Discussion

In this study, we evaluated the feasibility of T2 mapping at 0.55 T using ultra-fast spin-echo sequence. We used a physical phantom and spectrometer reference values to assess quantitative accuracy of T2 estimates, specifically addressing spin-spin relaxation using an ultra-fast spin echo sequence tailored for fetal MRI. In vitro results reveal that T2 times can be estimated with high spatial resolution using a total scan time of 16.5 min. Yet, HASTE sequence introduces a systematic bias and overestimates T2 times. In vivo, we show robust and repeatable T2 estimation for WM and GM on a healthy adult cohort.

Despite the expected reduction in SNR at 0.55 T, the high-resolution T2 maps generated from trilinear-interpolated T2w volumes across different TEs are comparable in quality to those at 1.5 T. The fast protocol, combined with multi-orientation redundancy and total variation denoising, enables effective reconstruction of high-quality T2 maps under restricted motion.

**Gaussian fitting can reliably estimate short T2 times but overestimates intermediate to larger T2 times due to spin contamination**

The in vitro results are somewhat mitigated. Although T2 estimation remains consistent across coils for low to intermediate T2 values, there is a persistent overestimation of T2 when using the HASTE sequence regardless of the fitting method. Only ROIs with T2 values within the range of 80–120 ms appear to be estimated fairly. These findings suggest that while the Gaussian model yields stable trends, its application to HASTE data leads to a systematic inflation of T2 values, particularly at longer relaxation times.

This likely reflects the lower signal-to-noise ratio and increased noise propagation at longer T2s, which can bias the fitting. Given that fetal and neonatal tissues generally have higher T2 values than adult brain tissue at the same field strength, this source of variation must be considered and studied in future in vivo applications. Nevertheless, for the typical adult brain T2 range at 0.55 T (approximately 80–150 ms (34)), the method performs fairly within expected variation. In this range, we expect accurate estimation of WM T2 and a slight overestimation of GM T2.

Interestingly, the same measurements at 1.5 T showed even larger deviations from spectrometer references, suggesting low-field MRI at 0.55 T may provide more faithful T2 estimates in this context — an advantage for clinical and research applications at lower field strengths.

A possible reason for T2 overestimation could be the use of the HASTE sequence design, which employs GRAPPA acceleration and has a partial Fourier factor that changes around 150 ms, potentially affecting the accuracy of the decay curve. Moreover, in the HASTE sequence, the center of the K-space is acquired at the effective echo time specified in the sequence parameters, but small delays occur for non-center regions. As a result, the T2 value we fit might be an apparent T2 value, requiring in-depth understanding of the HASTE sequence parameters for accurate fitting. This issue was especially evident in the NIST phantom analysis. A more sophisticated fitting approach would be necessary to accurately attribute the T2 times solely to tissue relaxation. According to (48), fitting under the assumption that noise follows a non-central Chi-square distribution could yield more precise results when GRAPPA acceleration is used, although this would drastically increase the model's complexity and computational cost.

## Rician noise component helps reduce bias but harms stability and map homogeneity

Qualitatively, the Gaussian noise model produced the smoothest and most homogeneous T2 maps. Although the Rician fit better matched the expected trend overall, it showed increased map noise and inhomogeneity, greater variability, especially at higher concentrations, whereas the Gaussian fit was more stable and reproducible across the full range. It must be noted that higher tolerance was used compared to the Gaussian fitting (Table S1) due to convergence issues encountered when fitting over the full volume, which includes regions of deionized water. This difference in tolerance settings may affect the stability of the resulting parameter maps. This demonstrates that while more complex noise assumptions reduce overestimation, they tend to be less stable and to over-correct, especially for tissues with shorter T2 relaxation times. These findings suggest a trade-off between accuracy relative to reference standards and spatial stability of the voxel-wise maps and is worth studying more deeply in subsequent studies.

## High resolution T2 mapping can be acquired using a 16.5 minute protocol

Acquiring 9 low-resolution volumes (3 TEs, 3 orientations per TE, each in ~1 minute 45 seconds) keeps the total acquisition time to 16.5 minutes and produces a high-resolution 1 mm³ volume within a 360x360 field of view, meeting clinical requirements for fetal MRI (15). We further show that coil-related variability is minimal for short and intermediate T2 times. Although T2 overestimation is slightly more pronounced with the soft abdominal coil for higher T2 times, our tissues of interest fall in the lower range, minimizing this effect. This limits domain shift between in utero and postnatal imaging, supporting smooth transitions from fetal to newborn imaging and improving pre- and postpartum care (49). Finally, the ablation study on TE coverage shows that, ideally, echo times should be uniformly distributed and span the full range of T2 times of interest to ensure accurate fitting across tissues. Future studies should validate if this T2 mapping framework is suitable for various pediatric imaging scenarios, including both newborn and fetal imaging.

## T2 mapping at 0.55T is coherent, robust and repeatable in vivo

The in vivo study on 10 healthy adult volunteers showed coherent T2 estimations for both WM and GM tissue types. The broader T2 distribution for GM with extended upper tails is likely due to partial volume contamination in regions adjacent to cerebrospinal fluid. The mean T2 times are consistent with those reported by (34), reaching 90 ms for WM and 120 ms for GM, measured on a 1.5 T scanner adapted for use at 0.55 T. The T2 maps delineate WM tracts, and the reduced SNR is mitigated by acquiring data in three different orientations, which enhances the final image quality.

The repeatability study further demonstrates stable T2 estimation, with the coefficient of variation (CoV) showing a consistent increase in inter-session variability compared to inter-run variability. As expected, the inter-subject CoV for our single-compartment T2 estimates remains acceptable in WM at below 5%, although slightly higher than values reported in previous studies for intra-/extra-cellular T2, which showed CoV below 2% for both WM and GM at 1.5 T (47). In contrast, our GM variability was notably higher at 17.7%, likely due to partial volume effects and the single-compartment model's limited ability to resolve mixed tissue signals.

The regression analysis of the average T2 times across specific ROIs showed a strong correlation, with the Pearson correlation coefficient confirming a highly significant relationship given two runs, sessions and subjects, further demonstrating the quantitative robustness of T2 estimation across different subjects using ultra-fast spin echo sequence.

## Limitations

The primary limitation of the repeatability study is the small cohort size of N = 10. The chosen recruitment pool included only healthy adult volunteers from one study center. A larger cohort will enable statistically more robust analysis. In addition, we conducted a single-site study using one scanner from a single vendor, which limits the generalizability of our findings. A multi-center study is necessary to validate the robustness of T2 mapping in domain shift settings compared to T2-weighted sequences, to assess the generalizability of T2 mapping at 0.55 T.

Additionally, this study only focused on T2 mapping without motion, where image registration and interpolation were enough to reconstruct a high-resolution 3D volume. However, the impact of more significant motion needs to be investigated and is especially relevant for fetal and pediatric cohorts, as this would require, in addition to the ultra-fast acquisition protocol, the use of super-resolution reconstruction algorithms which alter the absolute intensity values of T2-weighted images and thus would add new challenges for T2 mapping. Recent emerging approaches have begun to address this challenge, with (50) proposing a physics-informed super-resolution algorithm and reporting the first in-vivo fetal brain T2 maps at 0.55 T.

## Conclusion

This study introduces a simple, fast, and robust high-resolution T2 mapping imaging protocol at 0.55 T tailored to fetal MRI acquisitions, using an ultra-fast spin-echo multi-echo T2-weighted sequence combined with offline T2 fitting. We report T2 times of the healthy adult brain population at 0.55 T and a relatively short total acquisition time of 16.5 minutes, and we demonstrate that T2 mapping at low field is efficient and reliable, in vivo settings.

## References (max. 50)

## **Supplementary Materials**

**Table S1:** Fitting parameters

| Field strength | noise assumption | Initial condition (min bound, max bound) | | | L-BFGS-B Solver options | |
|---|---|---|---|---|---|---|
| | | **T2** | **M0** | **sigma** | **ftol** | **maxls** |
| **1.5 T** | **Gaussian** | 165 (10,2000) | 890 (signal(minTE), 10000) | N/A | 1E-06 | 50 |
| **0.55 T** | **Gaussian** | 165 (10,2000) | 650 (signal(minTE),10000) | | | |
| | **Gaussian-Rician & Rician** | | | 40 [2,200] | 1E-02 | 50 |

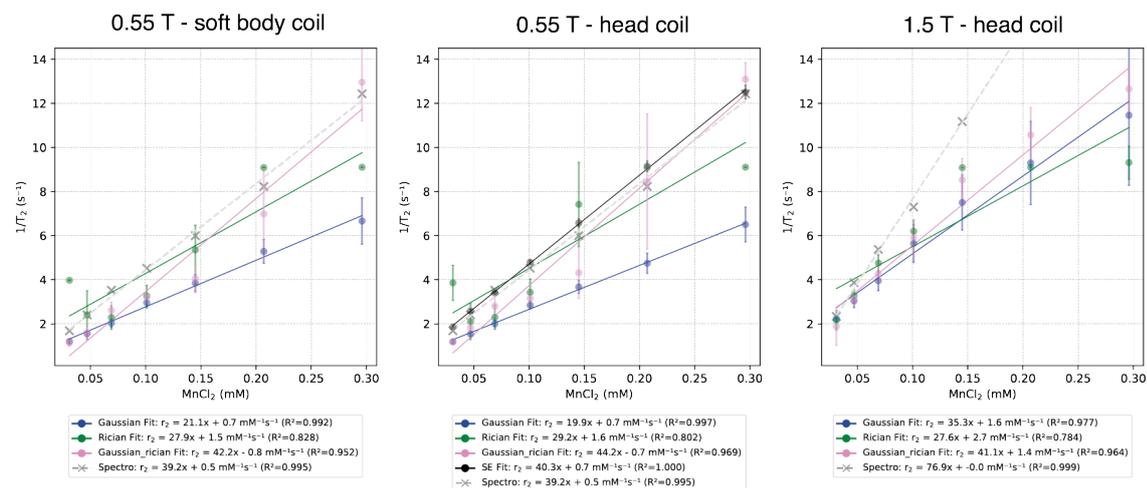

**Figure S1 : Relaxation rates 1/T2 versus MnCl2 concentration.** Shown are mean values and standard deviations in each region-of-interest. R2 are calculated from linear regression slopes. Gaussian, Gaussian-rician, Rician models are compared to SE and Spectrometer references values.

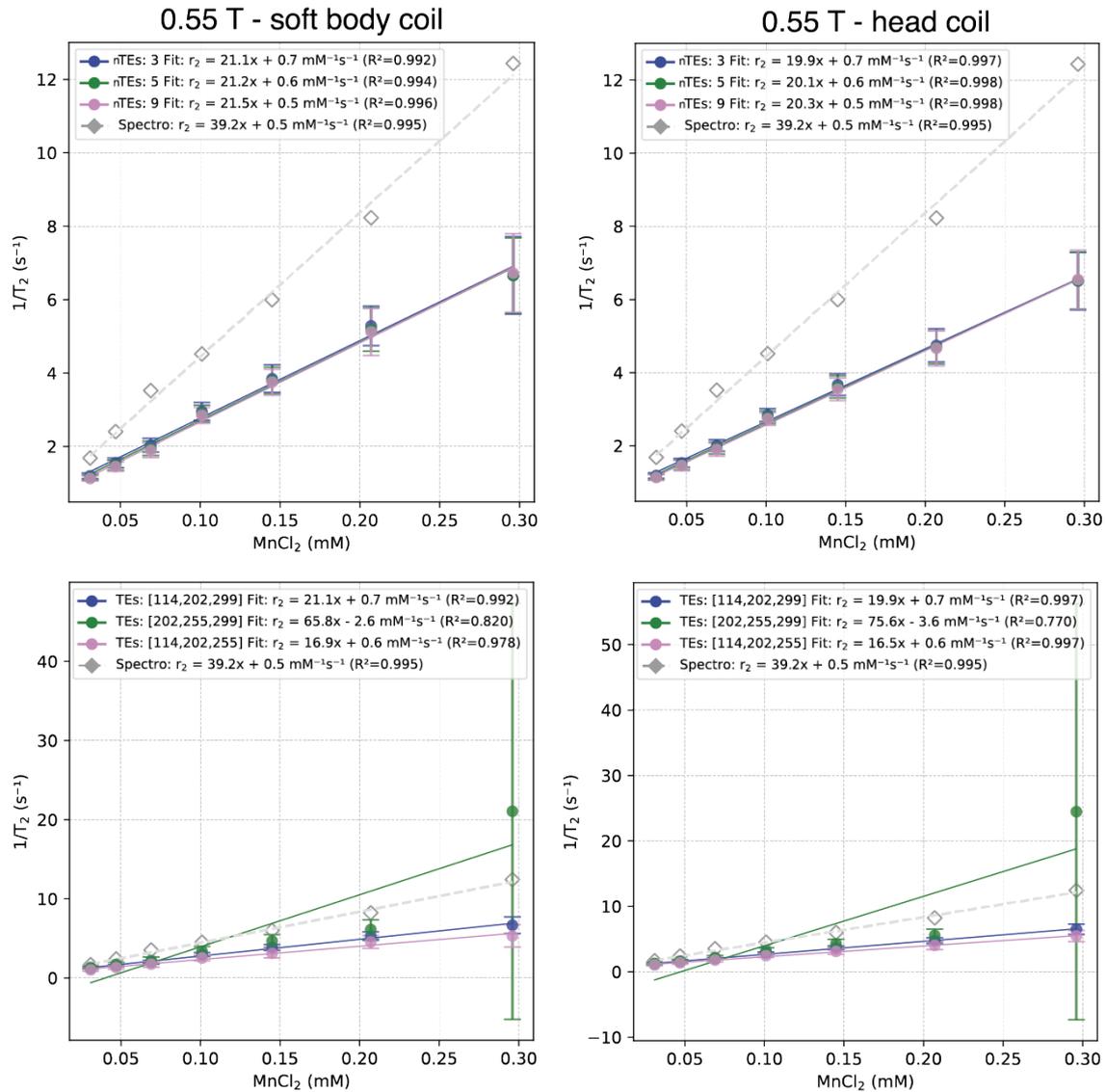

**Figure S2: Ablation Study on echo times.** Relaxation rates 1/T2 versus MnCl2 concentration. Shown are mean values and standard deviations in each region-of-interest. R2 are calculated from linear regression slopes.

**Table S2:** In vivo repeatability study CoV (%)

|  | CoV (%) | |
| --- | --- | --- |
|  | **White Matter** | **Grey Matter** |
| **inter-run** | $0.46 \pm 0.41$ | $1.49 \pm 1.30$ |
| **inter-ses** | $0.84 \pm 0.71$ | $1.43 \pm 1.42$ |
| **inter-sub 0.55 T** | $5.24 \pm 2.85$ | $17.7 \pm 10.3$ |
| **inter-sub 1.5 T** | $5.00 \pm 2.16$ | $12.51 \pm 4.45$ |

**Table S3:** CoV (%) Wilcoxon signed-rank test with Bonferroni correction

**Wilcoxon signed-rank test, with Bonferroni correction**

| | White matter ($n_{ROI}$=27) | | | Grey matter ($n_{ROI}$=41) | | |
|---|---|---|---|---|---|---|
| | T-statistic | p-value[*] | Decision | T-statistic | p-value[*] | Decision |
| **run vs ses** | 75 | 0.005 | Do not reject H0 | 399 | 0.69 | Do not reject H0 |
| **ses vs sub** | 0 | 1.49E-08 | Reject H0 | 0 | 9.09E-13 | Reject H0 |
| **0.55 vs 1.5** | 186 | 0.95 | Do not reject H0 | 285 | 0.060 | Do not reject H0 |

[*]Note: Significance level adjusted for multiple comparisons using Bonferroni correction ($\alpha = 0.05$ / number of ROI).

**Table S4:** Selected White Matter ROIs from ICBM-DTI-81 atlas

| |
|---|
| Pontine crossing tract (a part of MCP) |
| Genu of corpus callosum |
| Body of corpus callosum |
| Cerebral peduncle L+R |
| Anterior limb of internal capsule R |
| Anterior limb of internal capsule L |
| Posterior limb of internal capsule R |
| Posterior limb of internal capsule L |
| Retrolenticular part of internal capsule R |
| Retrolenticular part of internal capsule L |
| Anterior corona radiata R |
| Anterior corona radiata L |
| Superior corona radiata R |
| Superior corona radiata L |
| Posterior corona radiata R |
| Posterior corona radiata L |
| Posterior thalamic radiation (include optic radiation) R |
| Posterior thalamic radiation (include optic radiation) L |
| Sagittal stratum (include inferior longitudinal fasciculus and inferior fronto-occipital fasciculus) R |
| Sagittal stratum (include inferior longitudinal fasciculus and inferior fronto-occipital fasciculus) L |
| External capsule R |
| External capsule L |
| Cingulum (cingulate gyrus) L+R |
| Fornix (cres) / Stria terminalis (can not be resolved with current resolution) L |
| Superior longitudinal fasciculus L+R |
| Superior fronto-occipital fasciculus (could be a part of anterior internal capsule) L |
| Inferior fronto-occipital fasciculus L+R |

**Table S5:** Selected Grey Matter ROIs from Harvard-Oxford Cortical atlas

| | |
|---|---|
| Frontal Pole | Intracalcarine Cortex |
| Insular Cortex | Frontal Medial Cortex |
| Superior Frontal Gyrus | Juxtapositional Lobule Cortex (formerly Supplementary Motor Cortex) |
| Middle Frontal Gyrus | |
| Inferior Frontal Gyrus, pars triangularis | Subcallosal Cortex |
| Inferior Frontal Gyrus, pars opercularis | Paracingulate Gyrus |
| Precentral Gyrus | Cingulate Gyrus, anterior division |
| Temporal Pole | Cingulate Gyrus, posterior division |
| Middle Temporal Gyrus, anterior division | Precuneous Cortex |
| Middle Temporal Gyrus, posterior division | Cuneal Cortex |
| Middle Temporal Gyrus, temporooccipital part | Frontal Orbital Cortex |
| Inferior Temporal Gyrus, anterior division | Parahippocampal Gyrus, anterior division |
| Inferior Temporal Gyrus, posterior division | Parahippocampal Gyrus, posterior division |
| Inferior Temporal Gyrus, temporooccipital part | Lingual Gyrus |
| Postcentral Gyrus | Temporal Fusiform Cortex, anterior division |
| Superior Parietal Lobule | Temporal Fusiform Cortex, posterior division |
| Supramarginal Gyrus, anterior division | Temporal Occipital Fusiform Cortex |
| Supramarginal Gyrus, posterior division | Occipital Fusiform Gyrus |
| Angular Gyrus | Central Opercular Cortex |
| Lateral Occipital Cortex, superior division | Parietal Operculum Cortex |
| Lateral Occipital Cortex, inferior division | Occipital Pole |

**Table S6:** WM and GM T2 relaxation times across subject

| T2 (ms) | 0.55 T | | 1.5 T | |
|---|---|---|---|---|
| | WM | GM | WM | GM |
| **sub-002** | 117 ± 28 | 161 ± 35 | 113 ± 22 | 161 ± 45 |
| **sub-003** | 116 ± 15 | 167 ± 36 | 110 ± 16 | 156 ± 35 |
| **sub-004** | 115 ± 16 | 175 ± 39 | 110 ± 17 | 170 ± 46 |
| **sub-005** | 116 ± 17 | 176 ± 48 | 112 ± 18 | 165 ± 48 |
| **sub-006** | 122 ± 37 | 255 ± 155 | 113 ± 28 | 216 ± 80 |
| **sub-007** | 121 ± 39 | 247 ± 150 | 109 ± 24 | 181 ± 56 |
| **sub-008** | 117 ± 13 | 169 ± 49 | 111 ± 15 | 152 ± 36 |
| **sub-009** | 117 ± 14 | 174 ± 39 | 111 ± 15 | 174 ± 48 |
| **sub-010** | 121 ± 15 | 185 ± 72 | 113 ± 14 | 165 ± 45 |
| **sub-011** | 117 ± 16 | 170 ± 43 | 111 ± 17 | 166 ± 48 |
| **Mean across subjects** | **118** | **188** | **111** | **171** |

**Table S7:** Pearson correlation across field strength

| T2 (ms) | WM | | GM | |
|---|---|---|---|---|
| | **R** | **p-value** | **R** | **p-value**[+] |
| **sub-002** | 0.92 | 9.9e-12(***) | 0.59 | 5.7e-05 (**) |
| **sub-003** | 0.87 | 4.4e-09 (***) | 0.49 | 1.2e-3 (**) |
| **sub-004** | 0.84 | 5.5e-08 (***) | 0.15 | 0.34 (ns) |
| **sub-005** | 0.83 | 1.1e-07 (***) | 0.28 | 0.081 (ns) |
| **sub-006** | 0.72 | 2.9e-05 (***) | 0.13 | 0.41 (ns) |
| **sub-007** | 0.84 | 5.5e-08 (***) | 0.23 | 0.14 (ns) |
| **sub-008** | 0.82 | 1.8e-07 (***) | 0.21 | 0.19 (ns) |
| **sub-009** | 0.89 | 4.6e-10 (***) | 0.63 | 8.2e-06 (**) |
| **sub-010** | 0.77 | 3.0e-06 (***) | 0.18 | 0.27 (ns) |
| **sub-011** | 0.76 | 4.9e-06 (***) | 0.53 | 3.7e-04 (**) |
| **Mean across subjects** | **0.83** | | **0.34** | |

[+]Note: Significance level set at *** $\alpha = 0.001$/** $\alpha = 0.01$/* $\alpha = 0.05$, adjusted for multiple comparisons using Bonferroni correction ($\alpha$ = / number of ROI).

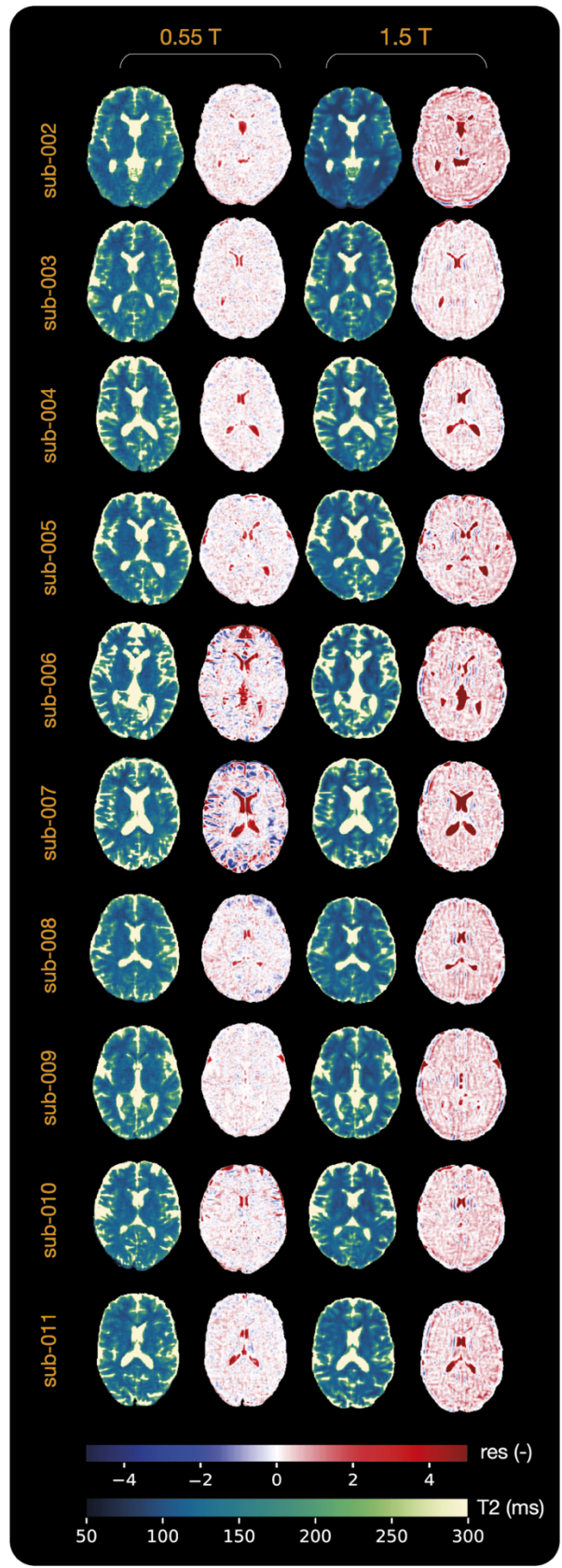

Figure S3: In vivo T2 maps and residuals of subjects at 0.55 T and 1.5 T